\documentclass[12pt,preprint]{emulateapj}

\usepackage{epsfig}
\usepackage{epstopdf}

\begin{document}

\def\beq{\begin{equation}}
\def\eeq{\end{equation}}

\newcommand{\kms}{\,{\rm km\,s^{-1}}}
\newcommand{\msun}{\, M_\odot}
\newcommand{\lsun}{\, L_\odot}
\newcommand{\lvsun}{\, L_{\odot,V}}
\newcommand{\mlrsun}{\Upsilon_{\odot,R}}
\newcommand{\mbh}{M_\bullet}
\newcommand{\mstar}{M_\star}
\newcommand{\mbulge}{M_{\rm bulge}}
\newcommand{\ml}{\Upsilon}
\newcommand{\gradml}{\alpha}
\newcommand{\mlr}{\Upsilon_R}
\newcommand{\mlavg}{\left< \Upsilon \right>}
\newcommand{\mlv}{M_\star/L_V}
\newcommand{\rinf}{\, r_{\rm inf}}
\newcommand{\reff}{\,r_{\rm eff}}
\newcommand{\scut}{\sigma_{\rm cut}}
\newcommand {\gtsim} {\ {\raise-.5ex\hbox{$\buildrel>\over\sim$}}\ }
\newcommand {\ltsim} {\ {\raise-.5ex\hbox{$\buildrel<\over\sim$}}\ }

\title{The Effect of Spatial Gradients in Stellar Mass-to-Light Ratio on Black Hole Mass Measurements}

\author{Nicholas J. McConnell \footnotemark[1], Shi-Fan S. Chen \footnotemark[2], Chung-Pei Ma \footnotemark[3], Jenny E. Greene \footnotemark[4], Tod R. Lauer \footnotemark[5], and Karl Gebhardt \footnotemark[6]}

\footnotetext[1]{Institute for Astronomy, University of Hawaii at M\={a}noa, Honolulu, HI; nmcc@ifa.hawaii.edu}
\footnotetext[2]{Phillips Exeter Academy, Exeter, NH}
\footnotetext[3]{Department of Astronomy, University of California at Berkeley, Berkeley, CA; cpma@berkeley.edu}
\footnotetext[4]{Department of Astrophysics, Princeton University, Princeton, NJ; jgreene@astro.princeton.edu}
\footnotetext[5]{National Optical Astronomy Observatory, Tucson, AZ; lauer@noao.edu}
\footnotetext[6]{Department of Astronomy, University of Texas at Austin, Austin, TX; gebhardt@astro.as.utexas.edu}

\begin{abstract}

We have tested the effect of spatial gradients in stellar mass-to-light ratio ($\ml$) on measurements of black hole masses ($\mbh$) derived from stellar orbit superposition models.  Such models construct a static gravitational potential for a galaxy and its central black hole, but typically assume spatially uniform $\ml$.  We have modeled three giant elliptical galaxies with gradients $\gradml \equiv d\log(\ml) / d\log(r)$ from $-0.2$ to $+0.1$.
Color and line strength gradients suggest mildly negative $\alpha$ in these galaxies. 
Introducing a negative (positive) gradient in $\ml$ increases (decreases)
the enclosed stellar mass near the center of the galaxy and leads to
systematically smaller (larger) $\mbh$ measurements.  For models with
$\alpha = -0.2$, the best-fit values of $\mbh$ are 28\%, 27\%, and 17\%
lower than the constant-$\ml$ case, in NGC 3842, NGC 6086, and NGC 7768,
respectively.  For $\alpha = +0.1$, $\mbh$ are 14\%, 22\%, and
17\% higher than the constant-$\ml$ case for the three respective galaxies.
For NGC 3842 and NGC 6086, this bias is comparable to the statistical
errors from individual modeling trials.  At larger radii, negative
(positive) gradients in $\ml$ cause the total stellar mass to decrease
(increase) and the dark matter fraction within one effective radius to
increase (decrease).
\\

\end{abstract}

\pagestyle{plain}
\pagenumbering{arabic}
%\addtolength{\topmargin}{-0.5in}

\maketitle

\section{Introduction}
\label{sec:intro}

Dynamical measurements of black hole masses ($\mbh$) at the centers of
nearby galaxies have opened new avenues for studying galaxy evolution, by
exposing correlations between $\mbh$ and various host galaxy properties.
Ongoing studies of these black hole scaling relations explore whether the
growth and quenching of black holes and galaxies are causally linked, or
only loosely connected by broad trends in cosmic evolution.  One way to
improve on the determination of the black hole scaling relations and the
cosmic scatter in $\mbh$ is to reduce uncertainties and systematic biases in
measurements of $\mbh$ in local galaxies.

Among the approximately 70 nearby galaxies with dynamically determined $\mbh$,
about 50 $\mbh$ have been measured using stellar kinematic data and orbit
superposition models (\citealt{mcconnellma13} and references therein).  A
number of assumptions made in the models can contribute to potential biases
in the inferred black hole masses.  Recent advances include modeling stellar mass and dark matter as separate components \citep[e.g.,][]{GT09}, and accounting for triaxiality \citep[e.g.,][]{vdB10}.

Another assumption that has been used in most previous orbit models
is that the stellar mass-to-light ratio, $\ml$, is constant throughout a
galaxy.  This assumption is clearly a simplification since spatial
gradients have been observed for many galaxy properties: e.g., color,
metallicity, age, and $\alpha$-element enhancement,
which can all be associated with spatial variations in $\ml$.
Early-type galaxies are often bluer towards their outer regions.
Decreasing metallicity outward is likely the dominant cause for the color
gradients in these galaxies \citep[e.g.,][]{Strom76, Tamura00}, with the
age gradient playing a minor or insignificant role
\citep[e.g.,][]{Saglia00, Tamura04}.  This picture is supported by recent
photometric studies of large statistical samples of early-type galaxies in
SDSS \citep{Tortora10a} as well as spectroscopic studies of individual
galaxies \citep{Spolaor09, Rawle10, Kuntschner10}.  The spatial coverage of
these studies is typically limited to the central regions of the galaxies
within one effective radius ($\reff$).  Measurements of ten early-type
galaxies have been extended to larger radii using deep observations with
integral field spectrographs \citep{Weijmans09, Greene12}.  The central
gradients in metallicity and absorption line strengths are seen to continue
to 2-4 $\reff$.  Gradients in color and inferred metallicity in galaxies'
globular cluster systems persist to $\sim 8 \reff$ \citep{Arnold11,Forbes11}.

Tight positive correlations between $\ml$ and color have been found in both
late-type and early-type galaxies \citep[e.g.,][]{Belldejong01, Bell03, Tortora11}, with redder colors corresponding to larger $\ml$.  By fitting synthetic spectral models to the observed optical color gradients,
\citet{Tortora11} have found mild negative gradients in $\ml$ for
local early-type galaxies with stellar ages $> 6$ Gyr, over a large range of
stellar masses ($\mstar \sim 10^{9.0}$-$10^{11.3} \msun$).  
Color and mass profiles have also been recently investigated in
galaxies at $0.5 < z < 2.5$ \citep{Szomoru13}.

In this letter, we assess the amount of systematic error in $\mbh$
resulting from the standard assumption of constant $\ml$ in the orbit
models.  We modify the orbit superposition code by \citet{Geb00b}
and introduce a spatial gradient in $\ml$ to the stellar component of the
potential.  We investigate the effects on $\mbh$ for three giant elliptical
galaxies that we have previously analyzed (assuming constant $\ml$): NGC
3842, NGC 6086, and NGC 7768 \citep{Mcconnell11a, Mcconnell11b,
  Mcconnell12}.  All three are massive ellipticals and brightest cluster
galaxies (BCGs) with high stellar mass $\mstar$ and velocity dispersion
$\sigma$: $\mstar = 1.55\times 10^{12}, 1.43\times 10^{12}, 1.16\times 10^{12}
\msun$ and $\sigma = 270, 318 , 257 \kms$ for NGC 3842, NGC 6086, NGC 7768,
respectively.  The black hole masses were predominantly constrained by
stellar kinematics from the integral-field spectrographs OSIRIS
\citep{Larkin} and GMOS \citep{ASmith02,Hook04}  Stellar kinematics at large radii were recorded with the Mitchell Spectrograph \citep{Hill} or adopted from the literature \citep{Carter,Loubser}.

A few prior studies of $\mbh$ have explored variations in $\ml$.  Recent
investigations of S0 galaxies NGC 1332, NGC 3368, and NGC 3489 modeled
bulge and disk components with separate $\ml$ values
\citep{Nowak10,Rusli11}.  \citet{Nowak07} assumed a separate $\ml$ value
for the nuclear disk in NGC 4486A.  \citet{Geb00b} assessed the $V-I$ color
gradients in NGC 3379 and applied the corresponding $\ml$ gradient to one
set of models; \citet{Capp02} applied a similar treatment to IC 1459.  For
both galaxies, models with a $\ml$ gradient and models with uniform $\ml$
yielded statistically consistent measurements of $\mbh$.  Here we extend
from prior investigations by examining systematic trends in measured $\mbh$
over a range of $\ml$ gradients.

\section{Stellar Orbit Models and $\ml$ Gradients}
\label{sec:models}
 
We use the axisymmetric orbit modeling algorithm of \citet{Geb00b,Geb03},
\citet{Thom05}, and \citet{Siopis}.
In this model, a galaxy is described by the density profile
\begin{equation}
  \rho(r,\theta) = \ml(r) \, \nu(r,\theta) +  M_\bullet \delta(r) + \rho_{\rm dm}(r) \,\; ,
\label{eq:rho}
\end{equation}
where $\nu(r,\theta)$ is the luminosity density derived from the galaxy's
deprojected surface brightness profile, and $\rho_{\rm dm}(r)$ represents
a spherical NFW or cored logarithmic (LOG) dark matter profile.
Equation~(\ref{eq:rho}) is used to construct a gravitational potential, and
kinematic models of the galaxy are constructed by propagating test
particles through the potential and computing their time-averaged
velocities (``orbits'') throughout a polar grid.  
Orbital weights are varied to fit the observed kinematics of the galaxy but are constrained such that the sum of weighted orbits exactly reproduces $\nu(r,\theta)$, which is decoupled from variations in $\ml(r)$.
The model velocity distributions are then compared to the observed kinematics,
yielding a goodness-of-fit statistic $\chi^2$.  
Each instance of the model 
adopts a single density profile and outputs a single $\chi^2$ value.  The
best-fit values and confidence limits for $\mbh$ and other parameters in
Equation~(\ref{eq:rho}) are determined by analyzing the distribution of
$\chi^2$ from many models.

%FIGURE -- color gradients
%
%\begin{figure}[!b]
\begin{figure}
\vspace{0.1in}
 \centering
 \hspace{-0.2in}
 \epsfig{figure=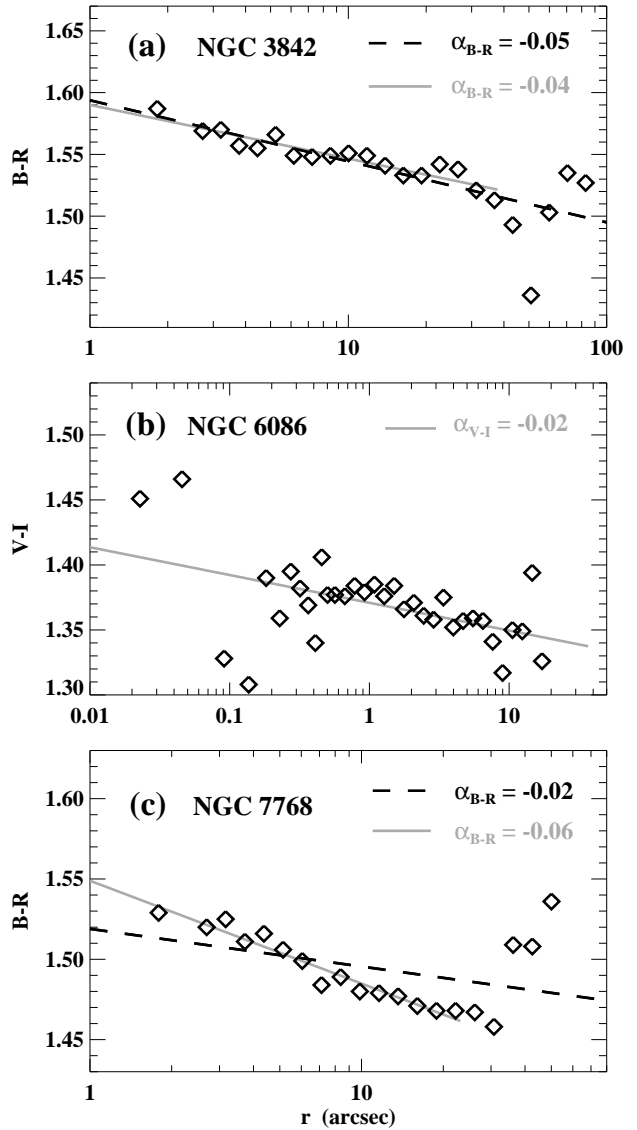,width=3.5in}
 \caption{Color vs. radius in three giant elliptical galaxies based on photometric data from the \textit{Hubble Space Telescope} and Kitt
Peak National Observatory.  In each panel, the black dashed line represents the best log-linear fit to all data points, and the solid grey line represents the best log-linear fit to data with $r \leq \reff$.  
For NGC 6086, our $V-I$ data extend to only $0.5\reff$ ($17''$).  $B-R$
data from \citet{Peletier90} extend to $1.2\reff$ and have a logarithmic
slope of $-0.06$.  }
\label{fig:colors}
\vspace{0.1in}
\end{figure}

Previous investigations using orbit superposition models typically have set
$\ml(r)$ to a uniform value, $\ml_0$.  To assess the impact of a gradient
in $\ml$, we set
\begin{equation}
  \alpha = \frac {d\log \ml}{d\log r}\, .
\label{eq:gradient}
\end{equation}
For each value of $\alpha$, we sample a two-dimensional grid of values for
$\mbh$ and the normalization of $\ml$.

The galaxies' color gradients indicate likely values of $\alpha$.
Figure~\ref{fig:colors} shows the color profile for each galaxy in our
sample.  The color gradients
$\delta$($B$-$R$)/$\delta(\log_{10}r)$ and
$\delta$($V$-$I$)/$\delta(\log_{10}r)$ range from $-0.02$ to $-0.06$.  
In Figure 2 of \citet{Tortora11}, the matching gradients in $\ml$ are $\approx
-0.1$ to 0 for SDSS galaxies.  Their Figure 3 shows the $\ml$ gradients as
a function of stellar mass and velocity dispersion and indicates 
different mean values of $\alpha$ for galaxies with different stellar ages.
Early-type galaxies with central age older than 6 Gyr have $\alpha \sim -0.2$ to $-0.1$ 
over a large range of $\sigma$,
but show a slight upturn toward $\alpha = 0$ at $\mstar > 10^{11} M_\odot$.  

For NGC 3842, we have an additional two hours of on-source data
from May 2012 observations with the Mitchell Spectrograph, allowing for a direct measurement of the
$\ml$ gradient.  As described in detail in \citet{Greene12}, we construct
elliptically averaged radial bins to maximize the S/N in the spectra. We
have a median S/N of 40 per pixel in the largest bin, at $r \approx 1.1 \reff$.  First,
weak emission features are iteratively corrected \citep{GF10} and velocity
dispersions are measured at each bin using {\tt pPXF} \citep{Cappellari04}.  Then
the stellar population properties are measured from the {\tt Lick\_EW+EZ\_Ages} algorithm
\citep{graves08}.  We use the $\alpha-$enhanced models of \citet{schiavon07}, assuming a Salpeter
IMF, to determine the luminosity-weighted $\ml$ at each radius.  
We find the resulting $\ml$ gradient to be $\alpha= -0.13 \pm 0.30$, broadly consistent with the color gradient in NGC 3842.  Uncertainties in the inner bins are dominated by systematic uncertainties in emission line and velocity dispersion measurements, while the outer bins include additional uncertainties from sky subtraction.

Gradients in the stellar initial mass function (IMF) are potentially an additional source of $\ml$ gradients within our galaxies.  The IMF varies with the [Mg/Fe] line ratio and the corresponding inferred timescale for star formation \citep{CvD12}.  Our line index measurements from Mitchell Spectrograph data of NGC 3842 are consistent with no radial gradient in [Mg/Fe].  
Therefore we do not expect IMF gradients to alter the likely range of $\alpha$.

Given the uncertainties discussed above, we have chosen to test $\alpha$ ranging from $-0.2$ to 0.1.
We employ $\alpha$ as a spherical, three-dimensional gradient in the models.  Integrating our model stellar mass profiles along the line of sight, we find the projected gradients to be $\approx 80\%$ of the three-dimensional $\alpha$.  The corresponding range of projected gradients, $-0.16$ to $0.08$, is appropriate with regards to the inferred gradients from photometry and spectroscopy.

%TABLE -- Trials
%
%\begin{table}[!b]
\begin{table}
%\vspace{0.1in}
\begin{center}
\caption{Results from Stellar Orbit Models}
\label{tab:models}
\begin{tabular}[b]{clcccc}  
\hline
 $\gradml$ & \multicolumn{1}{c}{$\mbh \;\;$} & $\ml(\rinf)$ & $\ml(\reff/10)$ & $\ml(\reff)$ & $\mlavg$ \\
& \multicolumn{1}{c}{($10^9 \msun$) $\;\;$} & ($\mlrsun$) & ($\mlrsun$) & ($\mlrsun$) & ($\mlrsun$)  \\[8pt]
\hline 
\multicolumn{6}{l}{NGC 3842 $\;$ ($\rinf = 1.2'' \;$ ; $\reff = 37.8''\;$ ; $D = 98.4$ Mpc)}  \\[8pt]
 -0.2 & $7.1^{+2.8}_{-2.8} \;\;$ & $6.4^{+0.9}_{-0.9}$ & $5.1^{+0.7}_{-0.7}$ & $3.2^{+0.4}_{-0.4}$ & $3.5^{+0.5}_{-0.5}$ \\[4pt]   %$4.0^{+0.6}_{-0.6}$ \\[4pt]
 -0.1 & $8.7^{+2.9}_{-2.9} \;\;$ & $5.8^{+0.8}_{-0.9}$ & $5.1^{+0.7}_{-0.8}$ & $4.1^{+0.6}_{-0.7}$ & $4.3^{+0.6}_{-0.7}$ \\[4pt]  %$4.6^{+0.6}_{-0.7}$ \\[4pt]
 -0.05 & $9.0^{+3.3}_{-2.5} \;\;$ & $5.5^{+0.7}_{-0.8}$ & $5.2^{+0.7}_{-0.8}$ & $4.6^{+0.6}_{-0.7}$ & $4.7^{+0.6}_{-0.7}$ \\[4pt]  %$4.9^{+0.6}_{-0.7}$ \\[4pt]
 0 & $9.8^{+2.9}_{-2.5} \;\;$ & $5.2^{+0.7}_{-0.7}$ & $5.2^{+0.7}_{-0.7}$ & $5.2^{+0.7}_{-0.7}$ & $5.2^{+0.7}_{-0.7}$ \\[4pt]  %$5.2^{+0.7}_{-0.7}$ \\[4pt]
 0.05 & $10.4^{+2.8}_{-2.5} \;\;$ & $4.8^{+0.7}_{-0.7}$ & $5.1^{+0.7}_{-0.7}$ & $5.7^{+0.8}_{-0.8}$ & $5.6^{+0.8}_{-0.8}$ \\[4pt]  %$5.4^{+0.8}_{-0.8}$ \\[4pt]
 0.1 & $11.2^{+2.6}_{-2.8} \;\;$ & $4.5^{+0.6}_{-0.6}$ & $5.0^{+0.7}_{-0.7}$ & $6.3^{+0.8}_{-0.9}$ & $6.2^{+0.8}_{-0.9}$ \\[4pt]  %$5.7^{+0.8}_{-0.8}$ \\[4pt]
 0.2 & $12.4^{+2.2}_{-2.6} \;\;$ & $3.9^{+0.5}_{-0.6}$ & $5.0^{+0.6}_{-0.7}$ & $7.8^{+0.9}_{-1.2}$ & $7.5^{+0.9}_{-1.1}$ \\[8pt]  %$6.4^{+0.7}_{-0.9}$ \\[4pt]
\hline
\multicolumn{6}{l}{NGC 6086 $\;$ ($\rinf = 0.24'' \;$ ; $\reff = 36.8''\;$ ; $D = 139.1$ Mpc)}  \\[8pt]
-0.2 & $2.6^{+1.5}_{-1.1} \;\;$ & $6.1^{+0.5}_{-0.5}$ & $3.6^{+0.3}_{-0.3}$ & $2.2^{+0.2}_{-0.2}$ & $2.8^{+0.2}_{-0.3}$ \\[4pt]  
-0.1 & $3.0^{+1.4}_{-1.1} \;\;$ & $5.3^{+0.4}_{-0.4}$ & $4.0^{+0.3}_{-0.3}$ & $3.2^{+0.3}_{-0.3}$ & $3.5^{+0.3}_{-0.3}$ \\[4pt] 
0 & $3.6^{+1.4}_{-1.2} \;\;$ & $4.4^{+0.4}_{-0.4}$ & $4.4^{+0.4}_{-0.4}$ & $4.4^{+0.4}_{-0.4}$ & $4.4^{+0.4}_{-0.4}$ \\[4pt]  
0.1 & $4.4^{+1.5}_{-1.3} \;\;$ & $3.6^{+0.2}_{-0.3}$ & $4.8^{+0.3}_{-0.4}$ & $6.0^{+0.4}_{-0.5}$ & $5.6^{+0.4}_{-0.5}$ \\[8pt]  
\hline
\multicolumn{6}{l}{NGC 7768 $\;$ ($\rinf = 0.14'' \;$ ; $\reff = 23.1''\;$ ; $D = 112.8$ Mpc)}  \\[8pt]
-0.2 & $1.0^{+0.5}_{-0.3} \;\;$ & $7.5^{+1.2}_{-1.2}$ & $4.3^{+0.7}_{-0.7}$ & $2.7^{+0.4}_{-0.4}$ & $2.7^{+0.4}_{-0.5}$ \\[4pt]  %$3.4^{+0.5}_{-0.5}$ \\[4pt]
-0.1 & $1.1^{+0.5}_{-0.3} \;\;$ & $6.3^{+0.8}_{-0.9}$ & $4.8^{+0.6}_{-0.7}$ & $3.8^{+0.5}_{-0.6}$ & $3.8^{+0.5}_{-0.6}$ \\[4pt]  %$4.3^{+0.5}_{-0.6}$ \\[4pt]
0 & $1.2^{+0.5}_{-0.3} \;\;$ & $5.0^{+0.6}_{-0.5}$ & $5.0^{+0.6}_{-0.5}$ & $5.0^{+0.6}_{-0.5}$ & $5.0^{+0.6}_{-0.5}$ \\[4pt]   %$5.0^{+0.6}_{-0.5}$ \\[4pt]
0.1 & $1.4^{+0.6}_{-0.3} \;\;$ & $4.0^{+0.4}_{-0.5}$ & $5.3^{+0.5}_{-0.6}$ & $6.7^{+0.7}_{-0.8}$ & $6.8^{+0.7}_{-0.8}$ \\[8pt]  %$6.1^{+0.6}_{-0.7}$ \\[4pt]
\hline   
\end{tabular}
\end{center}
Best-fit values and errors in $\mbh$ and $\ml$ are the median values and $68\%$ confidence limits derived from the cumulative likelihood method of \citet{Mcconnell11a}.  Stellar mass-to-light ratios $\ml$ are reported in $R$ band. $\mlavg$ is the average luminosity-weighted value of $\ml$ 
over the whole galaxy.
% between 0 and $\reff$  % -- use alternate values (commented) for last column.
\vspace{0.1in}
\end{table}

\section{Results}
\label{sec:results}

%FIGURE -- chi^2 contours
%
%\begin{figure}[!b]
\begin{figure}
 \centering
 \epsfig{figure=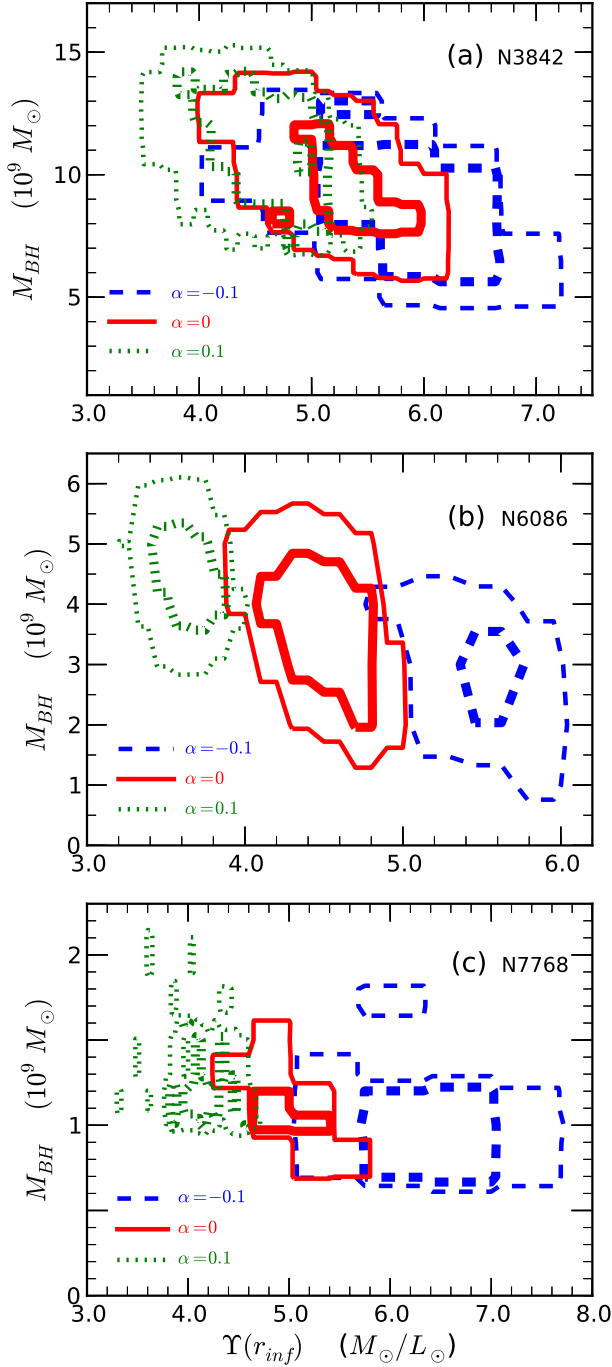,width=3.3in}
 \caption{Contours of $\chi^2$ as a function of
   $\ml(\rinf)$ and $\mbh$, for orbit superposition models with different slopes
   for the logarithmic $\ml$ gradient: $\alpha=-0.1$ (blue dashed), 0 (red solid), and $+0.1$
   (green dotted).  For each $\alpha$, the thick and thin contours represent $68\%$
   and $90\%$ confidence levels
   (corresponding to $\Delta \chi^2= 1.0$ and 2.71), respectively.
 }
\label{fig:contours}
%\vspace{0.1in}
\end{figure}

\begin{figure}
\vspace{-0.1in}
 \centering
 \epsfig{figure=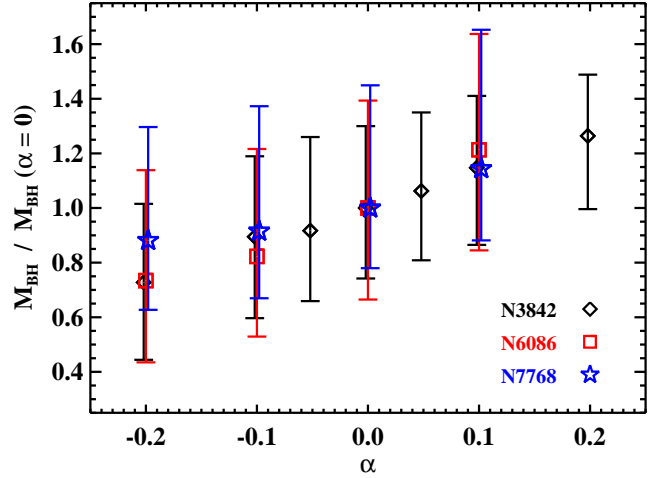,width=3.5in}
 \caption{Best-fit $\mbh$ vs. $\ml$ gradient $\alpha$ for models of NGC
   3842, NGC 6086, and NGC 7768.  The $y$-axis is the ratio of the best-fit
   $\mbh$ value relative to the best-fit value for models with uniform
   $\ml$ ($\alpha = 0$).  Error bars represent the statistical errors from individual trials (Table~\ref{tab:models}), also normalized relative to $\mbh(\alpha=0)$.}
\label{fig:BHalpha}
%\vspace{0.1in}
\end{figure}

Table~\ref{tab:models} and Figures~\ref{fig:contours} and \ref{fig:BHalpha} illustrate our best-fit
values of $\mbh$ and $\ml$ from modeling NGC 3842, NGC 6086, and NGC 7768
with different index $\alpha$ for the $\ml$ gradient.  
The stellar orbit models optimize the normalization of $\ml$, while the radial dependence is set by Equation~(\ref{eq:gradient}) for fixed $\alpha$.
Each galaxy's distance $D$, effective radius $\reff$, and black hole radius of influence, $\rinf \equiv
G\mbh/\sigma^2$, are listed in Table~\ref{tab:models}.  Since $\rinf$ is
much smaller than $\reff$ (with the ratio ranging from 0.006 to 0.03), we
list $\ml$ at three radii ($\rinf$, $0.1 \reff$, and $1 \reff$) for comparsion.
We also list the average luminosity-weighted stellar mass-to-light ratio
$\mlavg$, equal to the total stellar mass divided by the total luminosity.

\begin{figure}
%\vspace{-0.1in}
 \centering
 \epsfig{figure=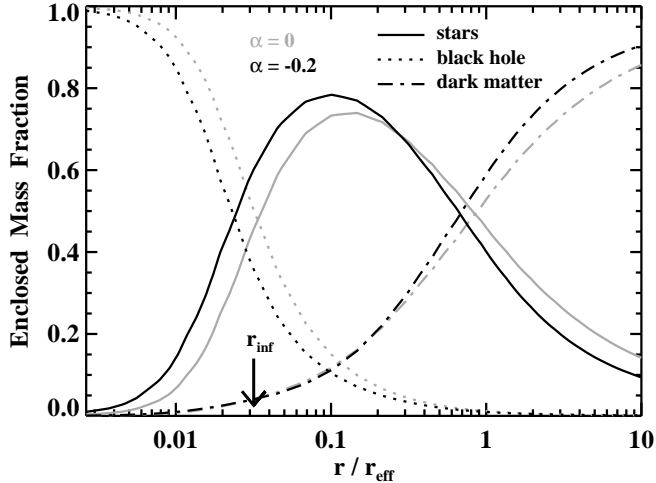,width=3.5in}
 \caption{Enclosed mass fractions vs. deprojected radius for stars, dark matter, and a
   black hole, in the best-fit gravitational potential for NGC 3842.  Two
   cases are shown: constant $\ml$ ($\alpha = 0$; grey) and a $\ml$ gradient
   with $\alpha = -0.2$ (black).  The black hole dominates the mass at
   small radii within the sphere of influence $\rinf$, stars dominate at
   intermediate radii of $r \sim 0.1 \reff$, and the dark matter halo takes
   over at $r\ga \reff$.}
\label{fig:fractions}
\vspace{0.1in}
\end{figure}

Table~\ref{tab:models} and Figure~\ref{fig:contours} show that within $\rinf$, there is a mild
degeneracy between stellar mass and black hole mass.
As $\alpha$
increases from negative toward positive values, $\ml(\rinf)$ decreases, causing the best-fit $\mbh$ to increase.
At large radii the trend with $\alpha$ is reversed, such that
the best-fit $\ml(\reff)$ and $\mlavg$ both increase as $\alpha$ increases.  The
intermediate radius of $\sim 0.1 \reff$ is roughly the pivot point at which
the best-fit $\ml$ has similar values for different $\alpha$.  
Our kinematic data for each galaxy fall mainly within $1 \reff$ and are
most thoroughly represented at small radii ($r \ltsim 3''$, or $\sim 0.1
\reff$), where the OSIRIS and GMOS integral-field spectrographs yield
measurements at multiple position angles.  This kinematic coverage drives
the normalization of $\ml$, such that best-fitting models for each value of
$\alpha$ have similar enclosed masses near $0.1 \reff$ for a given galaxy.
This is also the
scale on which stars dominate the total enclosed mass: much within $\sim 0.1
\reff$, the black hole begins to contribute, whereas the dark matter begins
to dominate at $\sim 0.5-2 \reff$ (Figure~\ref{fig:fractions}).

Figure~\ref{fig:BHalpha} shows a clear positive correlation between
$\alpha$ and the best-fit $\mbh$ value. As $\alpha$ decreases from 0 to
$-0.2$, the best-fit $\mbh$ decreases by $28\%$ in NGC 3842 and $27\%$ in NGC
6086.  For these galaxies, the bias in $\mbh$ over this range in $\alpha$
is comparable to the statistical uncertainty in $\mbh$ for each suite of
models with fixed $\alpha$.  NGC 7768 exhibits a slightly smaller bias of
$17\%$ in $\mbh$, versus statistical uncertainties $\sim 40\%$ from
individual trials.

For each galaxy, the value of $\chi^2$ for the best-fit model (for a given
$\alpha$) changes by less than 1.0 when $\alpha$ increases from $-0.2$ to
0.  By contrast, $\chi^2$ for the best-fit model increases by 2.5 as
$\alpha$ increases from 0 to $+0.2$ for NGC 3842, and $\chi^2$
increases by 1.0 and 2.8 for NGC 6086 and NGC 7768, respectively, as $\alpha$
increases from 0 to $+0.1$.  Models with a positive gradient in $\ml$ are
therefore mildly disfavored by our kinematic data.  This trend is
consistent with the range of $\alpha$ inferred from the three galaxies'
negative color gradients shown in Figure~\ref{fig:colors}.  

Figure~\ref{fig:fractions} shows the relative contributions of stars, dark
matter, and the black hole to the total enclosed mass as a function of deprojected
radius in NGC 3842.  Similar trends are seen for NGC 6086 and NGC 7768.
For simplicity, we have modeled a single dark matter halo for each galaxy,
for all values of $\alpha$.  As $\alpha$ decreases from 0 to $-0.2$, our
kinematic data prefer a model with more (less) stellar mass at 
$r \ll 0.1 \rinf$ ($r \gg 0.1 \rinf$).
Since we have assumed a fixed dark matter halo, the resulting enclosed stellar mass fraction is
decreased at large radii.  

If a range of dark matter profiles were sampled,
we would expect changes in $\ml(r)$ to alter the mass and/or shape of the
best-fitting dark matter halo.  Introducing a negative $\ml$ gradient would
redistribute stellar mass toward smaller radii, and the models could
reconstruct the outer mass profile by fitting a more massive halo,
amplifying the trend for dark matter in Figure~\ref{fig:fractions}.  Our models are are not sensitive to the inner slope of the dark matter profile, as the enclosed mass at small radii is dominated by stars and the black hole.

Since the dark matter halo and $\ml$ gradient both cause a non-uniform \textit{total} mass-to-light ratio, one might suspect that optimizing the dark matter profile for each value of $\alpha$ would decrease the variations in $\ml$ normalization and $\mbh$.  While our kinematic data for NGC 3842, NGC 6086, and NGC 7768 do not extend far enough to tightly constrain the dark matter profile, we have performed a simple test by varying $\alpha$ in models of NGC 3842 with no dark matter.  Even without dark matter, the overall trend of $\ml$ and $\mbh$ versus $\alpha$ is similar to those depicted in Figures~\ref{fig:contours}a and~\ref{fig:BHalpha}.

%FIGURE -- MBH vs alpha
%
%\begin{figure}[!t]
%

%FIGURE -- DM fraction vs alpha
%
%\begin{figure}[!t]

\section{Discussion}
\label{sec:disc}

We have run stellar orbit superposition models of three giant elliptical
galaxies, assuming a range of gradients defined by $\alpha = d\log(\ml) /
d\log(r)$ in the stellar mass-to-light ratio $\ml$.  We find that the
best-fit black hole mass $\mbh$ varies systematically with the strength and sign of $\alpha$.
As $\alpha$ decreases from 0 to $-0.2$, the best-fit values of $\mbh$
decrease by 28\%, 27\%, and 17\% for NGC 3842, NGC 6086, and NGC 7768,
respectively.  As $\alpha$ increases from 0 to $+0.1$, the best-fit values
of $\mbh$ increase by 14\%, 22\%, and 17\% for the three galaxies.
For comparison, individual trials yield statistical errors $\sim 30\%$-$40\%$ in $\mbh$.
Color and line strength gradients in these three galaxies suggest $\alpha \sim -0.2$ to $-0.1$. 

Our results suggest that gradients in $\ml$ may be a non-negligible source
of systematic error in measurements of $\mbh$ and $\mstar$.  The overall
effect of this error on the correlations between $\mbh$ and host galaxy
properties will depend on the distribution of $\ml$ gradients across the
sample of galaxies with existing $\mbh$ measurements.

All three galaxies in our study are BCGs with $\mstar >
10^{12} M_\odot$ and velocity dispersions $\sigma > 250 \kms$.
Extrapolating the trend illustrated in Figure~\ref{fig:BHalpha} to other values of
$\alpha$ will indicate the potential level of bias in dynamical
measurements of $\mbh$ that assume uniform $\ml$.  
A systematic variation
in $\alpha$ with galaxy stellar mass or velocity dispersion can have an
impact on the slopes of the $\mbh$-$M_{\rm bulge}$ and $\mbh$-$\sigma$
scaling relations, 
which are currently determined assuming $\alpha=0$ 
(e.g., \citealt{mcconnellma13} and references therein). 
For early-type galaxies, \citet{Tortora11} find a non-monotonic trend in the
dependence of $\alpha$ on $\mstar$ and $\sigma$, where $\alpha$ has
a minimum around  $\mstar \sim 1$-$3 \times 10^{10} \msun$
($\sigma \sim$ 100-160 $\kms$) and increases mildly toward the low- and high-$\mstar$ ($\sigma$) ends. The scatter, however, is large.  If $\alpha$ decreases (increases) with decreasing
$M_{\rm bulge}$ or $\sigma$, our results here indicate that the inferred
$\mbh$ would be systematically lower (higher) for smaller galaxies, thereby
steepening (flattening) the slopes of the scaling relations.  
In some galaxies where kinematic measurements superbly resolve $\rinf$, 
lower degeneracies between $\mbh$ and $\ml$ will reduce systematic errors in $\mbh$.

In addition to biasing measurements of $\mbh$, gradients in $\ml$ may
impact other attempts to decompose mass profiles into multiple components.
In particular, strong and weak lensing provide complementary data to
stellar kinematics for probing the total mass profiles of galaxies and
galaxy clusters.  In contrast to our small-scale kinematic data for NGC
3842, NGC 6086, and NGC 7768, lensing studies typically constrain the total
enclosed mass at larger radii of $\sim 1-100 \reff$.  In this case, the
inner slope of the dark matter profile can be degenerate with the stellar
mass profile.  Current methods estimate the stellar mass component using
stellar kinematics and/or population synthesis, under the assumption of
spatially uniform $\ml$
\citep[e.g.,][]{Jiang07,Sand08,Auger10, Tortora10b,Newman11,Newman12}.
Allowing for a gradient in $\ml$ in these studies could affect the
decomposition of the stellar and dark matter components.  For instance, in
the presence of a negative gradient in $\ml$, models sensitive to the
enclosed mass slope near $1 \reff$ might prefer a less cuspy dark matter
profile, to counteract the inward steepening in stellar mass.  

Gravitational lensing is better suited for probing lensing galaxies
at $z\sim 0.2$ and beyond, whereas
direct dynamical measurements of $\mbh$ are currently feasible
only in local galaxies within $\sim 150$ Mpc.
Other dynamical tracers such as globular clusters have been used to infer
galaxy mass profiles out to several effective radii
(e.g. \citealt{Romanowsky09, Murphy11}).  A joint analysis combining the
kinematics from our small-scale studies with those from the large-scale
tracers can provide more comprehensive measurements of the mass profiles of
stars and dark matter in early-type galaxies, from the black hole sphere of
influence out to several effective radii.

\medskip

This work was partially supported by NSF AST-1009663 and by a grant from
the Simons Foundation (\#224959 to CPM).  NJM is supported by the Beatrice
Watson Parrent Fellowship. All models were run using facilities at the
Texas Advanced Computing Center at the University of Texas at Austin.  We
thank Aaron Romanowsky for helpful discussions.

\end{document}